\documentclass{article}
\usepackage{spconf,amsmath,amssymb,amsfonts,graphicx}
\usepackage{cite}
\usepackage{algorithmic}
\usepackage{textcomp}
\usepackage{xcolor}
\usepackage{booktabs}
\usepackage{array}
\usepackage{multirow}

\title{Zero-shot sound event classification using a sound attribute vector with global and local feature learning}

\name{Yi-Han Lin, Xunquan Chen, Ryoichi Takashima, Tetsuya Takiguchi}
\address{Graduate School of System Informatics, Kobe University, Japan}

\begin{document}
\ninept

\maketitle

\begin{abstract}
This paper introduces a zero-shot sound event classification (ZS-SEC) method to identify sound events that have never occurred in training data.  In our previous work, we proposed a ZS-SEC method using sound attribute vectors (SAVs), where a deep neural network model infers attribute information that describes the sound of an event class instead of inferring its class label directly. Our previous method showed that it could classify unseen events to some extent; however, the accuracy for unseen events was far inferior to that for seen events. In this paper, we propose a new ZS-SEC method that can learn discriminative global features and local features simultaneously to enhance SAV-based ZS-SEC. In the proposed method, while the global features are learned in order to discriminate the event classes in the training data, the spectro-temporal local features are learned in order to regress the attribute information using attribute prototypes. The experimental results show that our proposed method can improve the accuracy of SAV-based ZS-SEC and can visualize the region in the spectrogram related to each attribute.

\end{abstract}


\section{Introduction}
\label{sec:intro}
Sound event classification (SEC) is a task in which we classify active sound events in a recording, such as the sound of running water, footsteps, or a moving car. 
It is expected to be applied in uses related to the care of the elderly and babies~\cite{wang2011robust,guyot2013elderly,Peng2009healthcare}, machine anomaly detection~\cite{pan2017cognitive,Koizumi2019,muller2020acoustic,koizumi2020description,Suefusa2020,dohi2021}, and so on. 
Recently, with the great strides made in the development of deep learning technology, it is becoming possible to analyze various sounds such as environmental sounds~\cite{chandrakala2019environmental,laffitte2016deep,imoto2020sed}. 
However, because a large amount of labeled data is required to train a deep learning model, 
it is a problem for SEC tasks that training data is difficult to come by for some events. 
For example, it is difficult to correct anomaly data in anomaly event detection because the given anomaly event rarely occurs. 

To overcome the problem of data scarcity, few-shot SEC methods have been proposed to classify sound events with only a few samples~\cite{chou2019learning,wang2020few,shi2020fewshot,yang2022mutual}. 
Chou \textit{et al.} propose an attentional similarity module to match transient sound events for few-shot SEC~\cite{chou2019learning}. 
In \cite{wang2020few} and \cite{shi2020fewshot}, prototypical networks were used to be an effective few-shot SEC method. 
Yang \textit{et al.} propose using mutual information loss to improve the class prototypes and feature extractor~\cite{yang2022mutual}. 
Although these few-shot SEC methods can achieve acceptable performance with only a few examples, they have only led to a classification process for predefined multiple classes. 
When encountering arbitrary sounds that might be unseen during training, the above-mentioned few-shot SEC methods have only rather limited classification capabilities.

Zero-shot learning (ZSL) extends the idea of few-shot classification by assuming that the labels we wish to predict at testing do not have available training data~\cite{rios-kavuluru-2018-shot}. 
ZSL has been studied widely in the field of computer vision (CV), and various methods have been proposed~\cite{wang2019,xian2016latent,huynh2020fine,sariyildiz2019gradient,Han2021} since the work carried out by Lampert \textit{et al.}~\cite{lampert2009}.
One of the representative approaches of ZSL in CV uses visual attributes~\cite{lampert2014,Stanislaw2014}, which describe the appearance of the class (e.g., horse shape, black and white, stripe, etc. for class ``zebra'') to classify images in the visual attribute space. 
In this way, even if a class has no training image data, this approach can identify the class through its attribute information. 

In contrast with many studies carried out on ZSL in CV, there has been little research carried out on ZSL for SEC.
Previous works~\cite{xie2021word,xie2021zero} propose methods of zero-shot SEC (ZS-SEC) using semantic embeddings. 
In those methods, a deep neural network (DNN) infers the semantic embedding representation from an input sound to classify the sound events in semantic space, and a word embedding generated by Word2Vec~\cite{mikolov2013distributed} from a class (event) label is used as a semantic embedding.
However, representing class information by word embeddings is considered inadequate for SEC because even though it reflects the semantic information of each word, it also contains much irrelevant information for the sound of each class. 
In our previous work~\cite{lin2022}, therefore, we proposed a sound attribute vector (SAV) that can directly describe the sound of the class as the visual attribute mentioned above describes the appearance of the class. 
The use of the SAV showed higher ZS-SEC accuracy than the use of word embeddings; however, the accuracy of classifying unseen events was still far behind that for classifying seen events.

In this paper, we propose a new ZS-SEC method that can learn discriminative global features and local features simultaneously to enhance the SAV-based ZS-SEC. 
In the proposed method, the SAV is inferred from an encoded input sound using two modules: a base module and a prototype module. 
The base module learns the discriminative global features of the input spectrogram to discriminate the attributes of each event class,
and the prototype module learns the spectro-temporal local features to regress the attributes of the target class.
In this way, the proposed method is expected to be able to enhance abilities of both discriminating event classes and inferring the attribute vector.
In addition, our proposed method can visualize the region in the spectrogram related to each attribute by calculating the similarity scores of the local features.
We confirm the effectiveness of the proposed method by comparing it to our previous SAV-based ZS-SEC method.

\section{Zero-shot learning using sound attribute vectors}
\label{sec:zsl_sav}

\begin{figure}[tb]
  \begin{center}
    \includegraphics[width=1\linewidth]{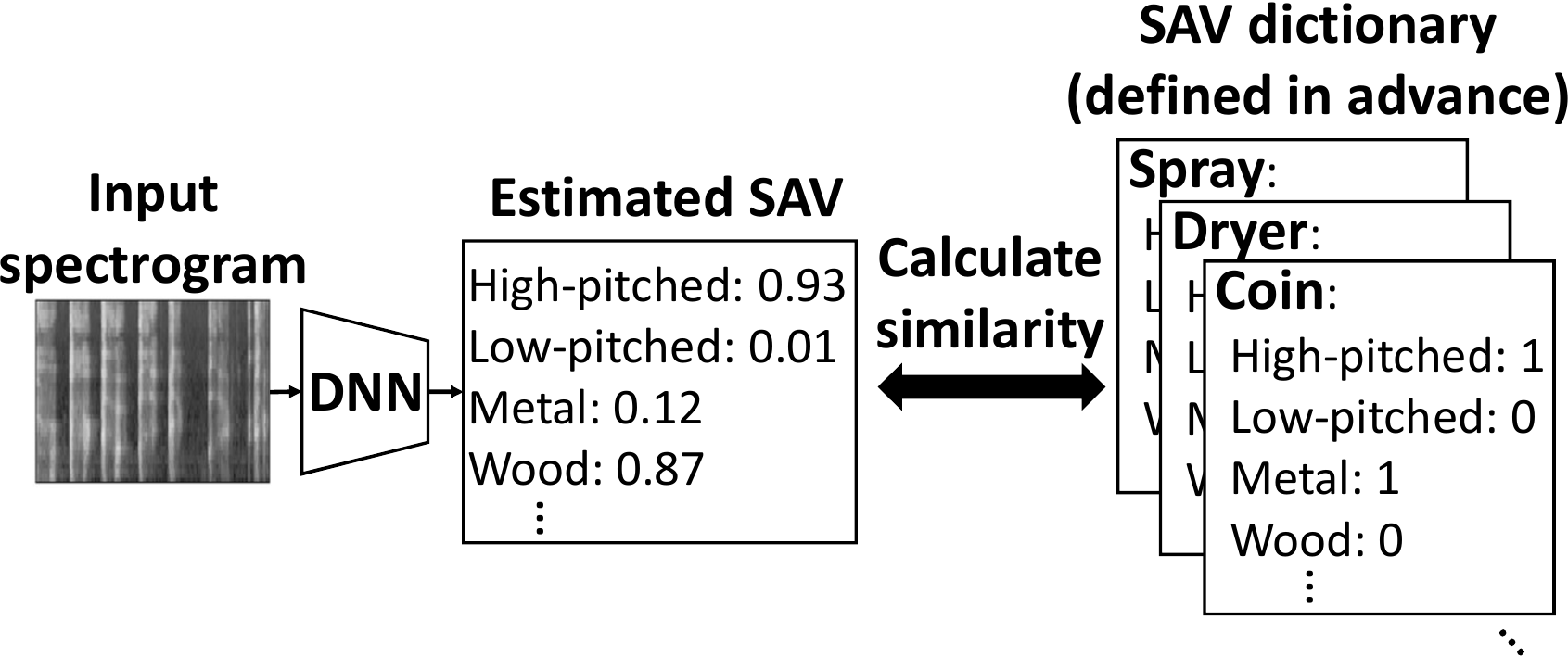}
  \end{center}
\vspace{-1.5em}
\caption{Overview of sound event classification based on sound attribute vectors.}
\label{fig:zsl_sav}
\end{figure}

Fig.~\ref{fig:zsl_sav} shows the overview of the SAV-based SEC system. 
In this system, the classification is performed in a semantic space defined by sound attributes instead of the class-label space, which is used in general SEC systems. 
The sound attributes describe the sound event and are expressed as a binary vector, in which ``1'' and ``0'' are assigned for corresponding and non-corresponding attributes, respectively. 
For example, for the event class ``falling coin'', attributes ``high-pitched'', ``metallic'', and ``collision'' will be set to 1, but ``low-pitched'' and ``wood'' will be set to 0.

The SAV-based system estimates the SAV of the input sound, and then the event class is identified by calculating the similarity between the estimated SAV and the SAV of each candidate in event classes defined in a dictionary in advance.
In this way, even if an arbitrary event has no training data, it can be identified if the SAV of that class is predefined.
In other words, even if we do not have the sound data of the event, we can identify it if we know what it sounds like.
In our previous work~\cite{lin2022}, we confirmed that the use of SAVs can provide higher ZS-SEC accuracy than the use of word embeddings. 
In this study, we define fifteen attributes for creating SAVs as shown in Table~\ref{tbl:attributes}.

\begin{table}[tb]
  \begin{center}
  \caption{Fifteen attributes defined in this study.}
  \label{tbl:attributes}
  \scalebox{1.0}[1.0]{
    \begin{tabular}{ l  l } \hline
      {\bf Descriptions}    & {\bf Attributes} \\ \hline \hline
      Sound pitch     &  {\it high-pitched}, {\it middle-pitched}, \\ 
                      &  {\it low-pitched}\\ \hline
      Sound length    & {\it long}, {\it middle}, {\it short} \\ \hline
      Material of     & {\it wood}, {\it metal}, \\
      sound source    & {\it plastic}, {\it ceramic} \\ \hline
      Other features  & {\it repeating} \\ &(repeating the 
                       same sound pattern), \\
                      & {\it noise-like} (white noise-like sound)    \\ \hline
      Situation of    & {\it falling} (something falling), \\
      the sound event & {\it collision} (something colliding), \\
                      & {\it many} (many sound sources)     \\ \hline
    \end{tabular}
  }
  \end{center}
\end{table}

\begin{figure*}[tb]
  \begin{center}
    \includegraphics[width=0.88\linewidth]{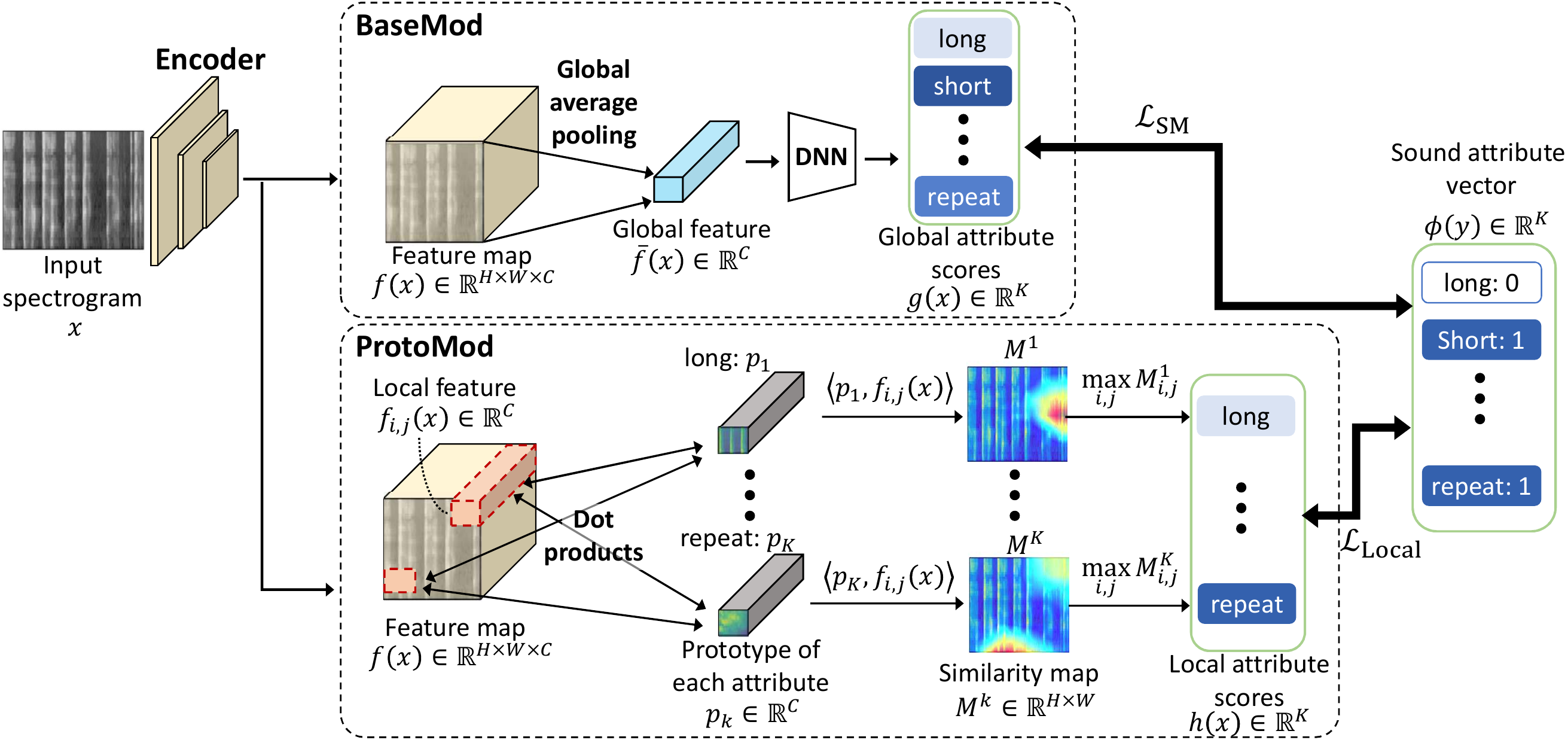}
  \end{center}
\vspace{-1.5em}
\caption{Overview of the proposed SAV-based SEC system.}
\label{fig:apn}
\end{figure*}

\section{Proposed Method}
\label{sec:method}

Fig.~\ref{fig:apn} depicts the overview of our proposed SAV-based SEC system. 
This system is based on the prototype attribute network proposed by Xu \textit{et al.}~\cite{xu2020attribute} for zero-shot image recognition, but some modifications have been made in our original system in order to apply that system to SEC (See section~\ref{sec:diff_apn}).
For a fixed-sized input spectrogram $x$, a feature map $f(x)\in \mathbb{R}^{H\times W\times C}$ is extracted by an encoder, where $H$, $W$, and $C$ denote the height, width, and channel of the feature map, respectively.
In this study, we employ VGGish~\cite{hershey2017cnn}, which is a model proposed for SEC, as the encoder.
Then, the extracted feature map is input to two modules: a base module (BaseMod) and a prototype module (ProtoMod) described in the next sections.

\subsection{Base Module (BaseMod)}
\label{sec:basemod}
BaseMod computes the global feature to infer the score of each attribute.
For the input feature map $f(x)$, the BaseMod calculates the global feature $\bar{f}(x)\in \mathbb{R}^{C}$ by applying global average pooling over the spatial axes (i.e., $\bar{f}(x) = \frac{1}{H\times W}\sum_{i=1}^{H}\sum_{j=1}^{W}f_{i,j}(x)$),
where $f_{i,j}(x)\in \mathbb{R}^{C}$ is extracted from the feature $f(x)$ at spatial location $(i,j)$.
Next, the global attribute scores $g(x) \in \mathbb{R}^{K}$, where $K$ denotes the number of attributes ($K=15$ in this study), are inferred by a DNN model (i.e., $g(x) = {\rm DNN}(\bar{f}(x))$).

The attribute scores can be scaled to the range $[0, 1]$ by applying sigmoid function. 
In our previous method~\cite{lin2022}, we trained the BaseMod and the encoder by using binary cross entropy (BCE) loss between the scaled attribute scores and the SAV of the teacher event class.
\begin{eqnarray}
\mathcal{L}_{\rm BCE} = {\rm BCELoss}({\rm Sigmoid}(g(x)), \phi(y)),
\label{eq:bce}
\end{eqnarray}
where $\phi(y) \in \mathbb{R}^{K}$ denotes the SAV defined for the event class $y$ corresponding to the input data $x$.

In the proposed method, we use a softmax loss instead of BCE loss in order to enhance the ability of the BaseMod and the encoder to discriminate the event classes:
\begin{eqnarray}
    \mathcal{L}_{\rm SM}=-\log \frac{\exp \left(\phi'(y)^{T} {\rm Tanh}(g(x)) \right)}{\sum_{y' \in \mathcal{Y}^{\rm seen}} \exp \left(\phi'(y')^{T} {\rm Tanh}(g(x))\right)},
\label{eq:softmax}
\end{eqnarray}
where $\mathcal{Y}^{\rm seen}$ is the set of event class labels in the training data.
$\phi'(y) = 2*\phi(y)-1$ and ${\rm Tanh}(g(x))$ are the SAV and the attribute scores scaled in $[-1, 1]$ instead of $[0, 1]$, respectively. Because we calculate the correlation between the SAV and the attribute scores in Eq.~(\ref{eq:softmax}), we apply this scaling in order to consider attributes that do not correspond to the event class for the correlation calculation.

\subsection{Prototype Module (ProtoMod)}
\label{sec:protomod}
Because our previous method uses only BaseMod with an encoder, it loses local information in the spectrogram.
However, it is considered that some attributes have a strong relationship to the location in the spectro-temporal space.
For example, the high-frequency component in the spectrogram may be important for the attribute ``{\it high-pitched}'', and the short-term region around the time a collision occurred may be important for the attribute ``{\it collision}''. 
For estimating those attributes, the important features are considered to be local, rather than the whole spectrogram.
For this motivation, we introduce a prototype module that learns local features that are important for estimating attributes.

ProtoMod has a learnable parameter $p_{k} \in \mathbb{R}^{C}$, named prototype, for each attribute $k$.
This prototype is initialized with random values and is trained as well as other parameters in whole model to represent a typical pattern of the local feature corresponding to each attribute. 
For an attribute $k$, we calculate the similarity between the prototype $p_{k}$ and the local feature $f_{i,j}(x) \in \mathbb{R}^{C}$ extracted from the feature $f(x)$ at spatial location $(i,j)$ by calculating their dot products.
By calculating the similarity for all locations, we obtain the similarity map $M_{i,j}^{k} = \langle p_{k}, f_{i,j}(x)\rangle$. 
We define the maximum value $\max_{i,j}M_{i,j}^{k}$ as the score of the attribute.
By defining the attribute scores as $h(x) = [\max_{i,j}M_{i,j}^{1}, \dots, \max_{i,j}M_{i,j}^{K}]$, we train the local features such that the estimated attribute scores match the SAV of the teacher event class by using mean square error loss.
\begin{eqnarray}
    \mathcal{L}_{\rm Local}=\|h(x)-\phi(y)\|_{2}^{2} 
\label{eq:local}
\end{eqnarray}
The total loss function considering both BaseMod and ProtoMod is defined as follows:
\begin{eqnarray}
    \mathcal{L}=\mathcal{L}_{\rm SM} + \lambda \mathcal{L}_{\rm Local},
\label{eq:loss}
\end{eqnarray}
where $\lambda$ denotes the weight for the $\mathcal{L}_{\rm Local}$.

\subsection{Classifying unseen events and two evaluation tasks}
\label{sec:inference}
When we classify the event class, we use the attribute scores output from the 
BaseMod\footnote{We have tried to use $h(x)$ from ProtoMod instead of $g(x)$ from BaseMod for the
classification; however, the performance degraded.}, 
and then 
we use Euclidean distance to calculate the similarity between the attribute scores and the SAV of each candidate of event class. 
Therefore, the event class is estimated as follows:
\begin{eqnarray}
    \hat{y} = \mathop{\rm arg min}_{y\in \mathcal{Y}^{\rm test}} \|{\rm Sigmoid}(g(x))-\phi(y)\|_{2}^{2},
\label{eq:sec}
\end{eqnarray}
where $\mathcal{Y}^{\rm test}$ is the set of event class labels in the test data. 
When the class labels of the test data consist only of unseen events (i.e.,  $\mathcal{Y}^{\rm test} = \mathcal{Y}^{\rm unseen}$), 
this evaluation task is referred to as a ZS-SEC task.
On the other hand, when the class labels include seen events (i.e., $\mathcal{Y}^{\rm test} = \mathcal{Y}^{\rm unseen} \cup \mathcal{Y}^{\rm seen}$), this evaluation task is referred to as a generalized ZS-SEC (GZS-SEC) task.

\subsection{Difference from the previous work in image recognition}
\label{sec:diff_apn}
As mentioned above, the idea of using BaseMod and ProtoMod has been proposed in computer vision~\cite{xu2020attribute}; however, 
some modifications have been made for our method to apply this approach to our ZS-SEC. 
1) We used VGGish as the encoder while ResNet101 was used in the original work. 
We preliminarily evaluated VGGish, VGG16/19, and ResNet50/101/152 as the encoder, and confirmed that the VGGish showed the best performance, but the training did not converge when ResNet50/101/152 were used.
2) We used an additional network behind the global average pooling layer in the BaseMod (``DNN'' in Fig.~\ref{fig:apn}), while a projection matrix was used in the original work, because we found it improved the SEC accuracy.
3) We scaled the outputs of the BaseMod and ProtoMod into the range $[0, 1]$ or $[-1, 1]$ depending on loss functions because we use binary values for expressing the attribute information. 
The original work, on the other hand, did not care about the scaling and used continuous values for expressing the attribute information.

\section{Experiment And Results}
\subsection{Experimental conditions}
\label{sec:condition}
We conducted our experiments on ZS-SEC and GZS-SEC tasks using the RWCP-SSD~\cite{nakamura2000acoustical} dataset, which involves various real-world environmental sounds.
From this dataset, we selected the following 6 classes of 543 sounds for unseen classes:
``bowl'' (striking a metal bowl with a metal rod), ``clock2'' (the sound of an electronic alarm clock), ``kara'' (shaking a rattle), ``maracas'' (shaking a maraca), ``ring'' (ringing a handbell), and ``tambouri'' (shaking a tambourine).
For the seen classes, we selected 62 classes of 5,647 sounds; however, we identified some classes (e.g., striking cherry wood and striking teak wood) as one class because those classes have similar sounds and the same SAVs. 
As a result, we redefined the 62 classes into 30 seen classes.
In the 30 seen classes, a total of 4,732 sounds were used for training the SEC model, and the other 915 sounds were used for testing.
For the 6 unseen classes, all 543 sounds were used for testing.

We used a 80-dimensional mel-spectrogram as the input features, and we fixed the number of frames to 100 (1 sec.) by cutting or zero-padding the original mel-spectrogram. Therefore, the size of the input spectrogram was fixed to $80 \times 100$.
We trained the VGGish as the encoder from scratch and did not use a pre-trained model.
The DNN behind the global average pooling layer has two middle layers, which consist of 4,096-nodes linear layers with ReLU activations, and the 15-nodes output layer.

\subsection{Results}
\label{sec:result}

\begin{table}[tb]
  \begin{center}
  \caption{Accuracy [\%] of classifying {\it unseen events} on a ZS-SEC task (accuracy for a random classifier: $1/6 = 16.7 \%$) and a GZS-SEC task (accuracy for a random classifier: $1/36 = 2.7 \%$).}
  \label{tbl:acc_zsl}
  \scalebox{1.0}[1.0]{
    \begin{tabular}{| l | c | c |} \hline
      Loss function  &  ZS-SEC & GZS-SEC\\ \hline \hline
      $\mathcal{L}_{\rm BCE}$ \cite{lin2022} &  61.0  & 1.3   \\ \hline
      $\mathcal{L}_{\rm BCE} + \mathcal{L}_{\rm Local}$ & 68.0 & 2.2     \\ \hline
      $\mathcal{L}_{\rm SM}$ &  64.6 & 0.7    \\ \hline
      $\mathcal{L}_{\rm SM} + \mathcal{L}_{\rm Local}$  &  {\bf 72.7} & 7.2     \\ \hline
    \end{tabular}
  }
  \end{center}
\end{table}

Table~\ref{tbl:acc_zsl} shows the classification accuracy for unseen classes on a ZS-SEC task and a GZS-SEC task with the variation of loss functions.
We set the weight parameter $\lambda$ to 1.0 for $\mathcal{L}_{\rm Local}$ with $\mathcal{L}_{\rm BCE}$, while we set it to 10.0 for $\mathcal{L}_{\rm Local}$ with $\mathcal{L}_{\rm SM}$, because the scale of value of $\mathcal{L}_{\rm SM}$ was about 10 times greater than $\mathcal{L}_{\rm BCE}$.
As shown in Table~\ref{tbl:acc_zsl}, both of discriminative loss function $\mathcal{L}_{\rm SM}$ and local feature training $\mathcal{L}_{\rm Local}$ improved the classification accuracy.
However, the accuracy severely degraded on the GZS-SEC task, showing lower accuracy than a random classifier (2.7\%) except for when using $\mathcal{L}_{\rm Local}$ with $\mathcal{L}_{\rm SM}$.
These results indicate that our system tends to mis-recognize unseen classes as seen classes.

\begin{table}[tb]
  \begin{center}
  \caption{Attribute detection performance.}
  \label{tbl:attr_est}
  \scalebox{1.0}[1.0]{
    \begin{tabular}{| l | c | c | c |} \hline
      Loss function  & Precision & Recall & F1score \\ \hline \hline
      $\mathcal{L}_{\rm BCE}$ \cite{lin2022} &  0.61 & 0.50 & 0.55     \\ \hline
      $\mathcal{L}_{\rm BCE} + \mathcal{L}_{\rm Local}$ & 0.64 & 0.54 & 0.59      \\ \hline
      $\mathcal{L}_{\rm SM}$ &  0.61 & 0.55 & 0.57     \\ \hline
      $\mathcal{L}_{\rm SM} + \mathcal{L}_{\rm Local}$  & {\bf0.65} & {\bf0.61} & {\bf0.63}     \\ \hline
    \end{tabular}
  }
  \end{center}
\end{table}

Table~\ref{tbl:attr_est} shows the performance of estimating the attribute vector. 
In this experiment, we detected the binary value of each attribute by using ${\rm Sigmoid}(g(x))$ with a threshold of 0.5 and evaluated the detection results by comparing them to the ground truth $\phi(y)$.
As shown in Table~\ref{tbl:attr_est}, our proposed method showed better estimation accuracy.

\begin{table}[tb]
  \begin{center}
  \caption{Comparison of model architectures for the encoder on a ZS-SEC task using loss function $\mathcal{L}_{\rm SM} +\mathcal{L}_{\rm Local}$. BaseMod and ProtoMod are not included in counting the number of layers and parameters. }
  \label{tbl:comp_enc}
  \scalebox{1.0}[1.0]{
    \begin{tabular}{| l | c | c | c |} \hline
      Encoder  &  \# layers & \# params. & Accuracy \\ \hline \hline
      VGGish & 6 & 4.5M  & {\bf 72.7}   \\ \hline
      VGG16 & 13 & 14.7M & 63.9    \\ \hline
      VGG19 & 16 & 20.0M & 58.0 \\ \hline
    \end{tabular}
  }
  \end{center}
\end{table}

We compared the VGGish to other model architectures as the encoder. 
As shown in Table~\ref{tbl:comp_enc}, VGGish showed the best performance in the compared models. 
We also evaluated ResNet50/101/152 because the previous work~\cite{xu2020attribute} used ResNet101; however, the model training did not converge when using ResNet models.

\begin{table}[tb]
  \begin{center}
  \caption{Classification accuracy (Acc.) [\%] of {\it seen events} (accuracy for a random classifier: $1/36 = 2.7 \%$) and attribute detection performance.}
  \label{tbl:acc_seen}
  \scalebox{1.0}[1.0]{
    \begin{tabular}{| l | c | c | c | c |} \hline
      Loss function  & Acc. & Precision & Recall & F1score  \\ \hline \hline
      $\mathcal{L}_{\rm BCE}$ \cite{lin2022} &  96.9 & 0.99 & 0.99 & 0.99      \\ \hline
      $\mathcal{L}_{\rm BCE} + \mathcal{L}_{\rm Local}$ & 98.1 & 0.99 & 0.99 & 0.99          \\ \hline
      $\mathcal{L}_{\rm SM}$ &  97.3 & 0.93 & 0.98 & 0.95                 \\ \hline
      $\mathcal{L}_{\rm SM} +\mathcal{L}_{\rm Local}$  &  97.3 &  0.97 & 0.98 & 0.98   \\ \hline
    \end{tabular}
  }
  \end{center}
\end{table}

Table~\ref{tbl:acc_seen} shows the classification accuracy and attribute detection performance for seen events.
As shown in Table~\ref{tbl:acc_seen}, our system showed high performance even when we use $\mathcal{L}_{\rm BCE}$ only. 
These results indicate that our system over-learns the sound of seen events, which is a possible reason for the severe degradation of the GZS-SEC task for unseen events in Table~\ref{tbl:acc_zsl}.

\begin{figure}[tb]
  \begin{center}
    \includegraphics[scale=0.45]{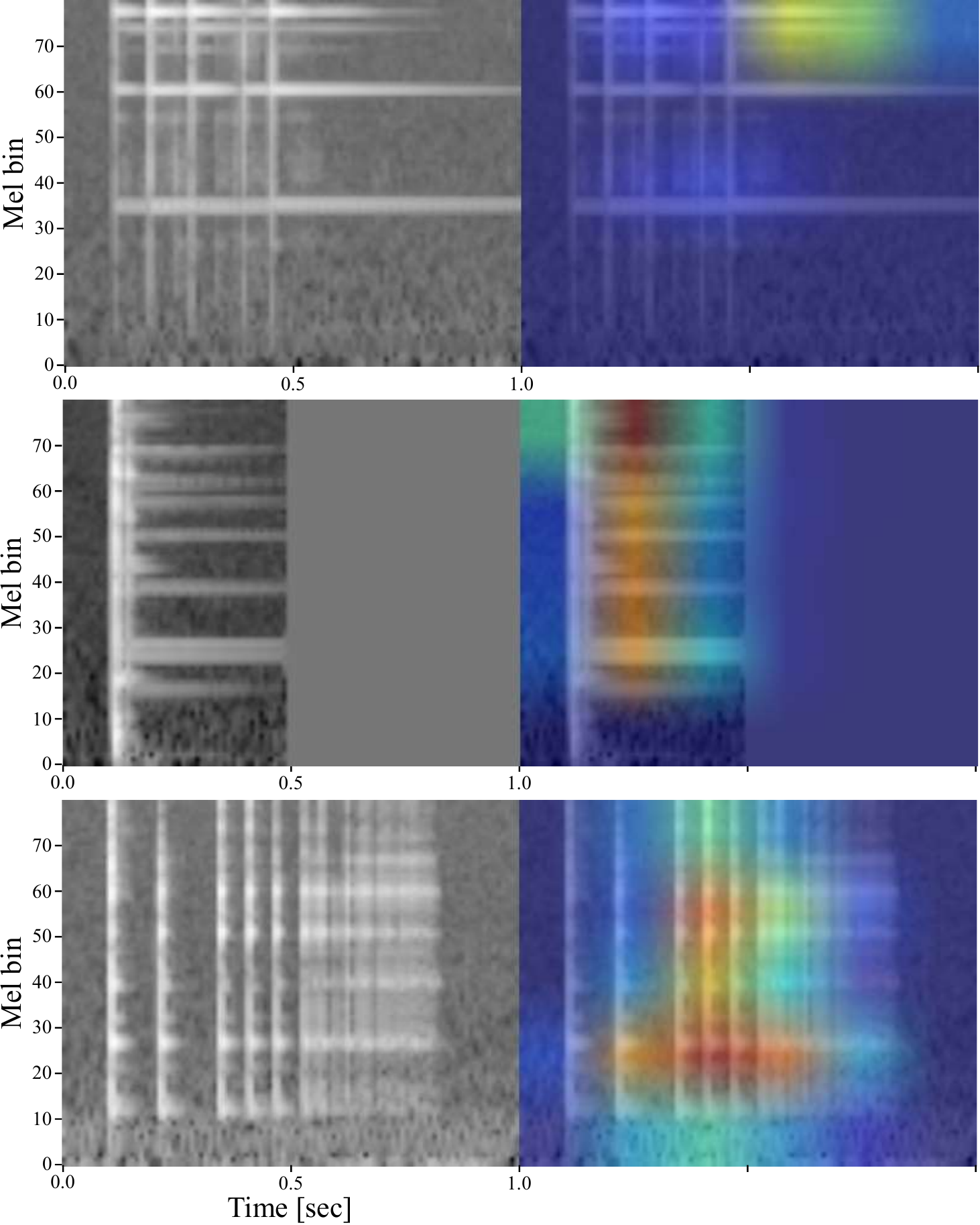}
  \end{center}
\vspace{-1.5em}
\caption{Input mel-spectrograms (left) and their similarity map $M^{k}$ for attributes $k$ calculated in ProtoMod (right); top: unseen event ``ring'' and its map for ``{\it high-pitched}'', middle: unseen event ``bowl'' and its map for ``{\it collision}'', and bottom: seen event ``dice2'' (a dice falls to the floor) and its map for ``{\it falling}''.}
\label{fig:heatmap}
\end{figure}

Fig.~\ref{fig:heatmap} shows the similarity map $M^{k}$ of the test data calculated in the ProtoMod.
When estimating the attribute ``{\it high-pitched}'' (top), the ProtoMod tends to focus on the region in which the high-frequency component is sparsely present.
When estimating the attribute ``{\it collision}'' (middle), the ProtoMod tends to focus on the time just after the collision occurs, not the moment it occurs.
When estimating the attribute ``{\it falling}'' (bottom), we can see that the ProtoMod focuses on the time region in which the dice is bouncing on the floor with successive collision sounds.
In this way, we confirmed that our proposed method can visualize the spectro-temporal region related to each attribute, which may help us to analyze the classification result.

\section{Conclusion}
\label{sec:conclusion}

In this work, we propose a SAV-based ZS-SEC method using global and local feature learning.
The experimental results show that our proposed method can significantly improve the ZS-SEC performance compared to the previous method. 
Furthermore, our proposed method can visualize the spectro-temporal region related to each attribute.
In our future work, we will investigate how to improve the accuracy of the GZS-SEC task.

\vfill\pagebreak
\clearpage

\bibliographystyle{IEEEbib}
\bibliography{refs}

\end{document}